\documentclass{ws-p8-50x6-00}

\def\btbb{\begin{tabbing}} \def\etbb{\end{tabbing}}
\def\beq{\begin{equation}} \def\eeq{\end{equation}}
\def\hs{\hskip} \def\la{\langle} \def\ra{\rangle} 
\def\hh{\hskip2pt} 
\begin{document}

\title{MULTIPARTICLE PRODUCTION IN NUCLEUS-NUCLEUS INTERACTIONS AT HIGH ENERGH}

\author{MUSTAFA NASR AND OMRAN TRIFEES}

\address{Department of Physics, Misurata University,\\
 Misurata-Libya P.O.BOX 1111}

\maketitle

\abstracts{We present and discuss results on multiplicity and angular 
distributions of shower particles produced in silicon-nucleus collisions at
4.5 A GeV/c. These results are compared with those obtained for different 
values of impact parameter. Shower width distribution has also been investigated.}

\section{Introduction}
Studies on nucleus-nucleus interactions at high energies have been attracting
more and more interest recently. It is believed [1-3] that these interactions 
may provide various information about the nuclear matter under extreme 
conditions of high temperature and densities. For this purpose, it is obviously
necessary to obtain a considerable amount of experimental information on 
diverse characteristics of such interactions; this applies particularly to the
multiple production of particles in the collisions of relativistic 
nucleus-nucleus collisions with relatively larger mass numbers of the incident
nuclei.

In the present work we present and discuss results on the characteristics and
angular distributions of relativistic singly charged particles produced in 
silicon-nucleus collisions at 4.5 A GeV/c. These results are compared with 
those involving carbon-nucleus collisions [4] at the same incident beam 
momentum.

\section{Experimental Detail}
The experimental data has been obtained by investigating a NIKFI-BR2 nuclear
photo-emulsion plates of dimension($16.9 \times 9.6 \times 0.06$) cm$^3$
irradiated by 4.5 A GeV/c silicon beam at Dubna Synchrophasotron. 1524 
interactions were collected by double, fast in the forward and slow in the 
backward direction, scanning of tracks under 15X eys-pieces and each 
interaction was looked at under 95X oil immersion objectives.

Secondary charged particles produced in each interaction are classified into
shower, grey and black tracks; the division of tracks into three distint
groups correspond to different physical processes through which they are 
envisaged to be produced. The shower tracks
have very large range and ionization less than 1.4{\sl $g_0$}, where 
{\sl $g_0$} represents the plateau density of a singly charged particles and
equals 29.6 grains per 100 micron. The grey tracks have ranges greater than
3 mm and their ionizations lie in the range 1.4{\sl $g_0$}-10{\sl $g_0$}.
The black tracks have ranges less than 3 mm and the ionization greater than
10{\sl $g_0$}. The number of shower, grey and black tracks produced in an
event are denoted by {\sl $N_s$}, {\sl $N_g$} and {\sl $N_b$} respectively.
Grey and black tracks together are referred to as the heavily ionizing 
particles or heavy tracks in an event and their number is represented by
{\sl $N_h (= N_g + N_b)$}.

Emission angles of all these tracks were measured by taking the space 
coordinates {\sl (x,y,z)} at three consecutive points on the same track and
at three other points on the incident beam track. Also, the space coordinates
{\sl (x$_0$,y$_0$,z$_0$)} of the production point were noted.

\section{Method of Target Identification}
It is quite difficult to identify the exact nature of the target because 
emulsion consists of H, C, N, O, Ag and Br nuclei. However, various methods 
have been tried by several authors [5-8] to identify the targets on the basis
of the distribution of heavily ionizing tracks, $N_h$. Usually, the events with
$N_h \leq 1$ are taken to be due to H targets; interactions characterized by
$2 \leq N_h \leq 6$ may be due to CNO and AgBr targets and the events with 
$N_h \geq 7$ are exclusively due to AgBr targets. This method of separation
of targets in the experiment is more accurate in hardron-nucleus interactions
because in that case the interactions are induced by very light nuclei. 
For the case of nucleus-nucleus interactions, this method is crude. The events
having $N_h > 8$ satify the condition fir AgBr, but for $N_h \leq 8$ events, 
there is an admixture of interactions due to CNO and peripheral collisions with 
AgBr targets. However, a method was suggested by Jakobsson and Kullberg[7] for 
the separation of AgBr events from the interactions having $N_h \leq 8$ by the
distribution of short-range tracks. This distribution reveals that there are
practically no tracks for the interactions with AgBr targets in the emulsion
having ranges between(10-50) $\mu$m in the interval of $N_h =2-8$. 
The events
due to AgBr nuclei with $N_h \leq 8$ and having at least one short track of
10 $\mu$m are due to recoil nuclei.

In this paper, an attempt has been made to separate the targets using the 
following criteria for the case of nucleus-nucleus interactions.

\btbb
Gay Ducat \hs12pt \= title of talks \kill
CNO events: \>$2 \leq N_h \leq 8$ and no track with range $\leq$ 10 $\mu$m\\
AgBr events: \>(i) $N_h > 8$ ,
            (ii) $N_h \leq 8$ {\bf and at least one track with range}\\
            \>  $\leq 10$ $\mu$m
           and no track with $10 \leq R \leq 50$ $\mu$m\\
H events:   \> $N_h \leq 1$, but do not fall in the above categories.\\        
\etbb
       
Table 1 gives the results using the above criteria along with the results of 
other similar efforts[4,9-11]. It may be seen from the table that the 
probability of events due to AgBr nuclei increases with increasing projectile
mass.

\vskip 2mm
{\footnotesize \hskip20pt
Table 1. Percentage of occurrence of interactions in nuclear emulsion.
\begin{center}
\begin{tabular}{ c c c c c c }
\hline
Momentum per & Projectile & H & CNO & AgBr & Ref. \\
Nucleon (GeV/c)& & & & & \\
\hline
3.0 & P & 18.00 & 49.50 & 32.50 & 9 \\
4.5 & $\alpha$ & 21.03 & 40.42 & 38.55 & 10 \\
4.5 & C & 21.29 & 30.87 & 47.84 & 4 \\
4.5 & Si & 15.29 & 33.79 & 50.92 & Persent work \\
1.8 & Fe & 23.13 & 22.64 & 54.23 & 11 \\
\hline
\end{tabular}
\end{center}}
\vskip3mm

{\footnotesize
Table 2. Mean  multiplicity and dispersion of relativistic charged particles
produced in carbon- (1st row) and silicon- (2nd row)
nucleus collisions at 4.5 A GeV/c.
\begin{center}
\begin{tabular}{cccccc}
\hline  
\multicolumn{2}{c}{H} &\multicolumn{2}{c}{CNO} &\multicolumn{2}{c}{AgBr}  \\
{$<N_s>$}& $D(N_s)$ & $<N_s>$ & $D(N_s)$ & $<N_s>$ & $D(N_s)$  \\
\hline
2.18 $\pm$ 0.18& 1.74 $\pm$ 0.13&5.04 $\pm$ 0.21& 
3.66 $\pm$ 0.15&8.92 $\pm$ 0.25&5.17 $\pm$ 0.18 \\
3.15 $\pm$ 0.19&2.98 $\pm$ 0.13&7.00 $\pm$ 0.26&
5.93 $\pm$ 0.18&16.30 $\pm$ 0.27& 10.95 $\pm$ 0.27\\
\hline
\end{tabular}
\end{center}}

\section{Experimental Results}
The values of mean multiplicities and dispersion of the relativistic charged 
particles produced in the interactions of silicon nuclei with emulsion nuclei
and the groups of events with $N_h \leq 1$, $2 \leq N_h \leq 8$ and 
$N_h >8$ are presented in Table 2. These results are compared with the values
obtained for carbon-nucleus interactions[4]. It is interesting to note that 
the mean multiplicity of the relativistic charged particles grows rapidly with
the increase in the mass of the projectile and target nuclei.

Figure 1 shows the multiplicity distributions of shower particles obtained for
the interactions of $^{28}$Si and $^{12}$C nuclei with emulsion for the 
three different targets. In Fig.2, we have plotted the multiplicity 
distributions of relativistic charged particles produced in $ ^{28}$Si-nucleus
collisions for different targets and these distributions 
are compared with $^{12}$C-nucleus collisions. It may be seen that tht 
distribution changes most rapidly
with increasing projectile and target mass; its broadening changes its shape.

In order to examine the dependence of mean shower particle multiplicity on 
target mass, the correlation between the mean multiplicity of shower and 
heavily ionzing particles has been investigated.

Figure 3 shows the correlation dependences $<N_s(N_h)>$ and $<N_h(N_s)>$ for
$^{28}$Si-nucleus interactions. We note that these dependences can be 
represented nicely by the following Linear relation with positive slopes.
   $$<N_s>=(0.81 \pm 0.02)_{N_h} + (2.05 \pm 0.52)\hfill (1)$$
   $$<N_s>=(0.75 \pm 0.02)_{N_s} + (1.88 \pm 0.58)\hfill (2)$$

For each group of $N_h$, the $N_h$ were accumulated into a composite star
and the forward F to backward B asymmetry of this star,
$A = (F -B)/ (F + B)$ was determined. Results for different $N_h$ groups are
shown in Fig. 4. From this figure, it is clear that the asymmetry A, which
is a measure of the four-momenta transfer to the target, is greater for smaller
$N_h$ values(peripheral collisions), whereas the results involving the whole 
data show relatively large momentums as well as "bounce off, side-splash'
effects in the case of non-peripheral collisions that characterizes the 
collective flow observed in heavy ion experiment.

The angular distribution of shower particles have been analyzed in terms of 
the rapidity parameter, which at high energies reduces to 
$\eta = -$ ln tan $\theta/2 $, 
where $\theta$ is the emission angle of a shower
particle with respect to the mean beam direction in the laboratory frame.
Figure 5 shows pseudorapidity distribution of relativistic charged particles
produced in $ ^{28}$Si-
nucleus interactions for the three different $N_h$ groups. From this 
figure, it is observed that the centroid of the $\eta$-distribution
shifts towards higher values of $\eta$ with increasing impact
parameter. Furthermore, the excess of particles arising due to the decreasing 
value of the impact parameter tends to appear only in the central region of 
the rapidity space.  

The variations of $<\eta>$ and rapidity dispersion D($\eta$) with $N_h$ are 
exhibited in Fig.6. It is interesting to note from the figure that $<\eta>$
decreases monotonically with increasing $N_h$ values. We notice that $<\eta>$
decreases slowly due to the loss of the  projectile energy with increasing
number of collisions. The energy is believed to be shared by more and more 
particles having relatively lower energies. This observation is incompatible
with the predictions of tube type models [12]. Furthermore, it may be noted
from this figure that there is no systematic variation in the rapidity 
dispersion \hh distribution \hh D($\eta$) \hh with \hh increasing \hh $N_h$, 
except in the region of very small values of $N_h$, where a large part of the 
cross-section is envisaged to be governed by the peripheral collision with 
probably one intranuclear nucleon.

The rapidity width (R) of an event is calculated using the following relation,
$R= \eta_{H} - \eta_{L}$,
where $\eta_{H}$ and $\eta_{L}$ are respectively the maximum and minimum 
pseudorapidity values in an interaction.  
In order to investigate the dependence of the rapidity width distribution on 
target size, the data have been divided into different $N_h$- intervals. The
shower width distributions for various $N_h$- intervals are displayed in Fig.7.
It may be noticed in this figure that the peaks of the R-distribution shift
towards the lower values of R with increasing impact parameter. This feature
can be explained in terms of the fact that shower particles produced with 
relatively larger angles would tend to appear in the target fragmentation 
region.

\section{Conclusion}
The following conclusions may be drawn from the results of the present work:

1- Multiplicity distribution of relativistic charged particles depends on both
projectile and target mass.

2- A linear dependence between the mean multiplicities of relativistic charged
and heavy particles is observed.

3- Angular distribution of shower particles depends strongly on the impact 
parameter.

\newpage
\vskip1cm
{\Large Figure Captions}
\vskip5mm

Fig.1 \hh Multiplicity distributions of shower particles for ($a$) silicon 

\hskip5mm
and ($b$) carbon
\vskip5mm

Fig.2 \hh Multiplicity distributions of relativistic particles for ($a$) H 

\hskip5mm
($b$) CNO ($c$) AgBr
\vskip5mm

Fig.3 \hh $\la N_s\ra$ vs. $N_h$ and $\la N_h\ra$ vs. $N_s$
\vskip5mm

Fig.4 \hh Asymmetry parameter vs. $N_h$
\vskip5mm

Fig.5 \hh Pseudorapidity distributions of relativistic particle produced 

\hskip5mm
in 4.5 GeV/$c$ Si interactions
\vskip5mm

Fig.6 \hh Dependence of $\la \eta\ra$ and $D(\eta)$ on $N_h$
\vskip5mm

Fig.7 \hh Shower width distributions in Si-nucleus interactions


\begin{thebibliography}{99}

\bibitem{1}H. H. Heckman, invited talk at Vth Int. Conf. on High Energy and
Nuclear Structure, Uppsala(1974).

\bibitem{2}W. Greiner, invited talk at Second Heavy-Ion Summer study, Berkeley,
1974 (LBL-3637, 1975),p.l.

\bibitem{3}K. D. Tolstov et al., preprint JINR, P1-8313, Dubna, Russia(1974).

\bibitem{4}T. Ahmad, Ph.D Thesis, Aligarh Muslim University, Aligarh India
(1991).

\bibitem{5}J. Babecki, Acta Phys. Pol. B6, 443(1975).

\bibitem{6}Z. V. Anzon et al., Sov. J. Nucl.Phys 22, 380(1976).

\bibitem{7}B. Jakobsson and R. Kullberg, Phys. Script 13, 327(1976).

\bibitem{8}M. Karabova et al., Sov. J. Nucl.Phys. 31,456(1980).

\bibitem{9}M. Bogdanski et al., Helv. Phys Acta 42, 485(1969).

\bibitem{10}DGKLMTW collaboration, JINR Dubna Commun., P1-8313(1974).

\bibitem{11}R. Bhanja et al., Nucl. Phys. A411, 507(1983).

\bibitem{12}Y. Afek et al., Topical Meeting on Multiparticle Production, Trieste
(1976). 

\end{thebibliography}
\end{document}